\documentclass[aps,preprint,showpacs,superscriptaddress,groupedaddress]{revtex4}  
\usepackage{graphicx}  
\usepackage{dcolumn}   
\usepackage{bm}        
\usepackage{amssymb}   
\usepackage{color}

\begin{document}



\title{Particle Acceleration during Magnetic 
Reconnection in a Low-beta Pair Plasma}
\author{Fan Guo}
\affiliation{Los Alamos National Laboratory, NM 87545 USA}
\author{Hui Li}
\affiliation{Los Alamos National Laboratory, NM 87545 USA}
\author{William Daughton}
\affiliation{Los Alamos National Laboratory, NM 87545 USA}
\author{Xiaocan Li}
\affiliation{Department of Space Sciences, University of Alabama in Huntsville, Huntsville, AL 35899, USA}
\author{Yi-Hsin Liu}
\affiliation{NASA Goddard Space Flight Center, Greenbelt, MD 20771, USA}

\begin{abstract}

\noindent Plasma energization through magnetic reconnection 
in the magnetically-dominated regime featured by low
plasma beta ($\beta = 8 \pi nkT_0/B^2 \ll 1$) and/or high 
magnetization ($\sigma = B^2/(4 \pi nmc^2) \gg 1$) is 
important in a series of astrophysical systems such as solar 
flares, pulsar wind nebula, and relativistic jets from black 
holes, etc. In this paper, we review the recent progress on 
kinetic simulations of this process and further discuss 
plasma dynamics and particle acceleration in a low-$\beta$ 
reconnection layer that consists of electron-positron pairs. 
We also examine the effect of different initial thermal 
temperatures on the resulting particle energy spectra. 
While earlier papers have concluded that the spectral 
index is smaller for higher $\sigma$, our simulations 
show that the spectral index approaches $p=1$ for 
sufficiently low plasma $\beta$, even if $\sigma \sim 1$.
Since this predicted spectral index in the idealized limit
is harder than most observations, it is important to consider effects that can lead to a softer spectrum such as open boundary simulations. We also remark that the effects
of 3D reconnection physics and turbulence on reconnection need to be addressed in the future.
\\

\end{abstract}

\maketitle


\section{Introduction}

Magnetic reconnection breaks and rejoins magnetic
field lines of force 
and reorders magnetic topology. Through this process
the magnetic energy is converted into plasma kinetic energy in bulk plasma flow, thermal and nonthermal particle distributions \citep{Biskamp2000,Priest2000}. Reconnection
plays a significant role in a wide range of laboratory, space, and astrophysical systems \citep{Zweibel2009,Yamada2010}.  
An important problem that remains unsolved is the acceleration of nonthermal charged particles in the reconnection region. While observations have shown strong evidence of 
particle acceleration associated with magnetic reconnection \citep{Oieroset2002,Masuda1994,Krucker2010}, 
the primary acceleration mechanism is still under debate 
\citep{Ambrosiano1988,Hoshino2001,Pino2005,Drake2006,Fu2006,Pritchett2006,Oka2010,Drury2012,Sironi2014,Dahlin2014,Guo2014,Guo2015,Zank2014,LeRoux2015}.  It is
worthwhile to point out that the 
acceleration mechanism may depend critically on how reconnection actually proceeds in large 3D system, a subject which is currently an active area of research. 
During the past decade, it has been shown by both two-dimensional
magnetohydrodynamic (MHD) simulations and particle-in-cell (PIC) kinetic simulations
that for a large system with weak collisions, the secondary tearing instability leads to
fractal reconnection layers with chains of plasmoids developed and the reconnection rate can be independent
of the Lundquist number \citep{Loureiro2007,Daughton2009,Bhattacharjee2009,Uzdensky2010}.
However, effects like MHD turbulence and 3D
physics that could be
important to the physics of magnetic reconnection have not been fully understood to reach a consensus \citep{Lazarian1999,Kowal2009,Loureiro2009,Daughton2011,Daughton2014,Guo2015,Huang2015b,Huang2015c}.

In astrophysical problems such as solar flares, pulsar wind nebula, 
and relativistic jets in gamma-ray bursts and active galactic nuclei, magnetic reconnection 
is often invoked to explain high-energy emissions from the strongly magnetized flows 
\citep{Thompson1994,Drenkhahn2002,Abdo2011,Tavani2011,Zhang2011,Uzdensky2011SSRv,
McKinney2012,Arons2012,Yu2013,Zhang2015b}. For 
relativistic plasmas, it is useful to define the 
magnetization parameter $\sigma \equiv B^2/(4\pi n m c^2)$, which indicates the ratio of the 
energy density of the magnetic field to the rest energy 
density of the plasma. For nonrelativistic plasmas,
it is more appropriate to use plasma beta 
$\beta = 8 \pi nkT/B^2$ that represents the ratio between
plasma thermal energy to magnetic energy.  In high-energy
astrophysics, it is often estimated that the magnetization
parameter can be much greater than unity $\sigma \gg 1$(or 
$\beta \ll 1$) and the Alfv$\acute{e}$n speed 
approaches the speed of light $v_A \sim c$.
To explain the observed high-energy emissions,
often an efficient mechanism from energies in the magnetized flow 
into nonthermal particles is required 
\citep[e.g.,][]{Celotti2008,Zhang2011}. 
 In the high-$\sigma$ regime, magnetic reconnection 
is the major candidate for 
converting magnetic energy and producing nonthermal particles and radiations. For a number of other systems 
such as solar corona and disk corona \citep{Tsuneta1997,Galeev1979}, although the Alfv$\acute{e}$n speed is not relativistic, the magnetic energy can greatly exceed the plasma thermal energy so $\beta \ll 1$. 
During magnetic reconnection a large fraction of the magnetic 
energy can be unleashed explosively into plasmas within a short time 
typically on the order of the Alfv$\acute{e}$n crossing time. 

Much of the recent progress on particle energization during reconnection has been made through first-principles 
kinetic simulations that self-consistently include the particle dynamics and the 
microphysics that is necessary to describe collisionless magnetic reconnection.  
While earlier numerical studies have identified multiple acceleration processes \citep{Hoshino2001,Drake2006,Fu2006,Pritchett2006,Oka2010,Dahlin2014}, 
recent simulations have revealed an efficient nonthermal 
acceleration that gives hard power-law like energy 
distributions \citep{Sironi2014,Guo2014,Melzani2014b,Werner2016,Li2015,Guo2015,Guo2016}. In this paper, we summarize the relevant 
progress in this area.  We also further study and clarify 
particle energization in the magnetically dominated plasmas 
with focuses on the regime with a low-$\beta$ pair plasma 
($\beta \ll 1$, $m_i = m_e$). We report new results on the influence of the initial plasma temperature on 
the hardness of the spectrum.
While earlier papers conclude that the spectral index is 
smaller for higher $\sigma$, our simulations show that the 
spectral index approaches $p=1$ for sufficiently low plasma 
$\beta$, even if $\sigma \sim 1$. The spectrum is harder 
than most of the observed energy spectra. This suggests that 
to explain the observed spectral index, it is important to 
consider effects that can lead to a softer spectrum such as
the effect of open boundaries.
We discuss recent progress in Section 2. The detailed 
numerical methods and parameters are presented in Section 3.
Section 4 discusses the main results of the paper. In 
Section 5, we summarize the results and outline several 
important problems to be addressed in the future.

\section{Nonthermal Particle Acceleration in Magnetic Reconnection Layers Revealed by 
Kinetic Simulations}

Earlier kinetic studies have identified numerous
different acceleration mechanisms in the reconnection layer. 
\citet{Hoshino2001} showed that several processes
can occur in a single reconnection layer -- in the X-line region \citep{Speiser1965,Huang2010} 
and along the separatrix region \citep{Egedal2012,Wang2014},
particles can get accelerated in the nonideal electric field
 and then further accelerated due to grad-B drift and the curvature drift in the magnetic 
pileup region \citep{Egedal2015}, where the electric field is mostly ideal $\textbf{E} = -\textbf{v}\times \textbf{B}/c$.
\citet{Drake2006} have further developed the Fermi
mechanism inside the magnetic islands as particles 
get bounced
at two ends of islands repeatedly \citep{Fu2006}. \citet{Oka2010} summarized a
number of basic acceleration mechanisms and concluded that 
island coalecensce region is an important acceleration site. In 
these regions the reconnected flux ropes interact and create 
new reconnection sites. For a large-scale reconnection layer 
that contains multiple X-points, the acceleration is more 
complicated and needs to be studied in a collective manner.
\citet{Dahlin2014,Guo2014} and 
\citet{Li2015} have shown that, for a large-scale kinetic
simulations that contain multiple X-regions, statistically 
the curvature 
drift acceleration along the reconnecting electric field is the 
dominant acceleration when the guide field is weak. The 
nonideal electric field only contributes to a small fraction of 
energy conversion in the simulation. The effect of a guide 
field that is normal to the reconnection plane can 
significantly alter the dominant acceleration mechanism
\citep{Pritchett2006,Fu2006,Huang2010}. It should be noted that 
in situ observations at the magnetotail have found evidence 
for those acceleration mechanisms. Although energetic particles
associated with diffusion regions have been discovered and 
detected by spacecraft observation \citep{Wu2015}, the flux ropes appear to be 
a stronger sources of energetic electrons \citep{Wang2010a,Wang2010b,Huang2012}. 
Betatron acceleration and Fermi acceleration are found to be 
important acceleration mechanism further away from the X-points
\citep{Hoshino2001,Fu2011,Fu2013,Wu2013,Huang2015a}.

Initial kinetic simulations of relativistic 
magnetic reconnection have found that strongly
nonthermal distributions can be
generated at the X-line region through direct acceleration 
in the diffusion region \citep{Zenitani2001}. While particles
get further accelerated in the magnetic pileup regions, the 
overall energy distribution in the whole domain does not show obvious power-law distributions \citep{Zenitani2007,Bessho2012,Cerutti2013}. Over the past few 
years, several groups have reported hard power-law
distributions $1 \leq p \leq 2$ when $\sigma \gg 1$
\citep{Sironi2014,Guo2014,Guo2015,Melzani2014b,Werner2016,Guo2016}. These new simulations found power-law distributions in the whole reconnection region, suggesting reconnection in 
magnetically-dominated regime may be a strong source of nonthermally energetic charged particles. While these results
appear to be repeatedly confirmed, the dominant acceleration
mechanism and the formation mechanism for the power-law distribution are still under debate. Through tracing
the guiding-center drift motions of particles in PIC simulations, \citet{Guo2014,Guo2015} have shown that the dominant 
acceleration mechanism is a first-order Fermi mechanism 
through curvature drift motions
of particles in the electric field induced by the reconnection 
generated flows. By considering an energy continuity equation,
it has been shown that a power-law distribution can be generated
when a continuous injection and Fermi acceleration $dE/dt = \alpha E$ are considered. The solution also gives a 
general condition for the formation of the
power-law particle energy distribution, i.e., the acceleration
time scale is shorter than the time scale
for particles injected into the reconnection region $\tau_{acc} < \tau_{inj}$. This mechanism gives rise to the formation 
of hard power-law spectra 
$f \propto (\gamma-1)^{-p}$ with spectral index
approaching $p = 1$ for a sufficiently high
$\sigma$ and a large system size. 
Following this work, the power-law distribution
has also been reported in nonrelativistic
reconnection simulation with a 
low-$\beta$ proton-electron plasma \citep{Li2015}, 
indicating the power-law distribution can develop 
in a larger parameter regime than previous 
expected high-$\sigma$ regime. In the simulations
with magnetically-dominated proton-electron plasmas, 
both electrons and protons develop significant power-law
distributions \citep{Guo2016}. 
On the other hand, \citet{Sironi2014} argued that the initial 
nonthermal energization at the X-line regions is crucial 
for the generation of the power-law distribution
\citep{Zenitani2001,Bessho2012}. In the vicinity of 
the X-lines, the 
initial distribution is 
energized into a nonthermal distribution even flatter 
than the overall distribution but with a limited energy 
range. This nonthermal distribution gets further 
accelerated in flux ropes to eventually develop into the 
observed spectra \citep{Spitkovsky2015}. \citet{Nalewajko2015}
have shown statistically that the acceleration in the island
merging region is a dominant source of nonthermal acceleration.
However, the analysis is mostly based on the acceleration site rather
than the acceleration mechanism.

\section{Numerical Methods}
Kinetic studies of magnetic reconnection have 
shown that current layers with thicknesses on the 
order of kinetic scales -- skin depth $d_i$ or thermal 
gyroradius $\rho_i$ -- are subject to reconnection.
We assume a situation where intense current sheets develop 
within a magnetically dominated plasma. This can be achieved 
through various processes such as striped wind geometry 
\citep{Coroniti1990,McKinney2012}, field-line foot-point 
motion \citep{Titov2003,Wan2014}, and turbulence cascade 
\citep{Zhdankin2012,Makwana2015}. During reconnection, the 
critical parameters that quantify the energization in the 
current layer are the magnetization parameter
$\sigma_e \equiv B^2/(4\pi n_e m_e c^2)$ and plasma beta 
$\beta_e \equiv 8 \pi n_ekT_e/B^2$. The numerical simulations
presented in this paper are initialized from a force-free 
current layer with $\textbf{B} = B_0 \text{tanh} (z/\lambda) \hat{x} + B_0 \text{sech} (z/\lambda) \hat{y}$
\citep{Che2011,Liu2013,Liu2015,Guo2014,Guo2015,Li2015}, 
corresponding to a magnetic field with magnitude $B_0$ 
rotating by $180^\circ$ across the central layer with
a half-thickness of $\lambda$. The initial distributions 
are Maxwellian with a spatially uniform density $n_0$ and 
thermal temperature $T_e = T_i$.
Particles in the central sheet have a net drift 
$\textbf{U}_i = - \textbf{U}_e$ to represent a current 
density $\textbf{J} = en_0(\textbf{U}_i - \textbf{U}_e)$ 
that is consistent with 
$\nabla \times \textbf{B} = 4\pi \textbf{J}/c$. Since the force-free
current sheet does not require a hot plasma component to balance the 
Lorentz force, this initial setup may be more suitable to study reconnection 
in low $\beta$ and/or high-$\sigma$ plasmas. We have also used relativistic
Harris current sheet \citep{Zenitani2007,Kirk2003} and found the two initial
setup generally gives similar results, although the hot plasma component in general results in a Maxwellian-like distribution that may dominate over the nonthermal 
distribution. 


In our present simulations, we assume plasma consists of electron-positron
pairs with mass ratio $m_i/m_e = 1$. No external guide field is included 
but there is an intrinsic guide field associated with the central sheet 
for the force-free setup. During the evolution the guide field will be 
expelled from the layer into the flux rope/island regions and later
the current sheet closely resembles antiparallel reconnection
\citep{Liu2015}. In this study, we vary the initial thermal temperature
to examine its influence on the resulting energy 
spectra. This has not been fully examined in previous papers. The full 
particle simulations are performed using the VPIC code \citep{Bowers2009}, 
which explicitly solve Maxwell equations and push particles in a 
relativistic manner. In the simulations, $\sigma$ is adjusted by changing 
the ratio of the electron gyrofrequency $\Omega_{ce} = eB/(m_e c)$ to the
electron plasma frequency $\omega_{pe} = \sqrt{4 \pi ne^2/m_e}$, 
$\sigma \equiv B^2/(4\pi n_e m_e c^2) = (\Omega_{ce}/
\omega_{pe})^2$. We primarily focus on 2D simulations with $\sigma = 1 \rightarrow 100$ and box sizes $L_x\times L_z =300 d_i \times 
150 d_i$, $600 d_i \times 300 d_i$, and $1200 d_i
\times 600 d_i$, where $d_i$ is the inertial length $c/\omega_{pe}$.
We also show a 3D simulation that discussed previously
\citep{Guo2014,Guo2015}. The 3D simulation has dimensions 
$L_x \times L_y \times L_z = 300 d_i \times 194 
d_i \times 300 d_i (N_x \times N_y \times N_z = 2048 \times 2048 \times 2048)$, $kT_e = kT_i = 0.36 m_e c^2$, and $\sigma=100$.
The half-thickness of the current sheet is $\lambda = 6 d_i$ for all cases.
For both 2D and 3D simulations, we have averagely more than 
$100$ electron-positron pairs in each cell. The boundary conditions for 2D 
simulations are periodic for both fields and particles in the $x$-direction, 
while in the $z$-direction the boundaries are conducting 
for the field and reflecting for the particles.
In the 3D simulations, the boundary conditions are 
periodic for both fields and particles in the 
$y$-direction, while the boundary conditions in the 
$x$ and $z$ directions are the same as the 2D cases.
A weak long-wavelength perturbation \citep{Birn2001} 
with $B_z = 0.03 B_0$ is included to 
initiate reconnection. 
All the simulations presented here 
show excellent energy conservation with violation of energy
conservation small enough to accurately
determine the particle energy spectra \citep[See the related discussion 
in][]{Guo2015}.

\section{Simulation results}
Figure $1$ shows the evolution of the current layer 
for force-free setup in two-dimensional simulations with 
$\sigma = 100$. For comparison, a 2D cut from a three-dimensional 
simulation from earlier studies \citep{Guo2014,Guo2015} is also presented. 
They show some common features for such 2D and 3D kinetic simulations 
of magnetic reconnection starting from a perturbation. For the 2D case,
the current sheet first thins down under the influence of the perturbation. 
The extended thin sheet then breaks into many fast-moving
secondary plasmoids due to the growth of the secondary tearing 
instability. These plasmoids coalensce with each other and eventually merge into a single 
island on the order of the system size. In the pair plasma case,
it has been shown that this secondary tearing instability and 
plasmoids facilitate fast reconnection and energy release
\citep{Daughton2007,Guo2014,Guo2015}. The 3D simulations show
 that the kink instability develops and interacts with the
 tearing mode, leading to a turbulent reconnection layer \citep{Yin2008,Guo2014,Guo2015}. 
 It has been shown that
 although the strong 3D effects can modify the current layer, 
 small-scale flux-rope-like structures with 
 intense current density develop repeatedly as a result of the 
 secondary tearing instability \citep{Guo2015}. The reconnection
 rate is roughly the same for the two cases \citep{Guo2015}.
 
\begin{figure}
\begin{center}
\includegraphics[width=\textwidth]{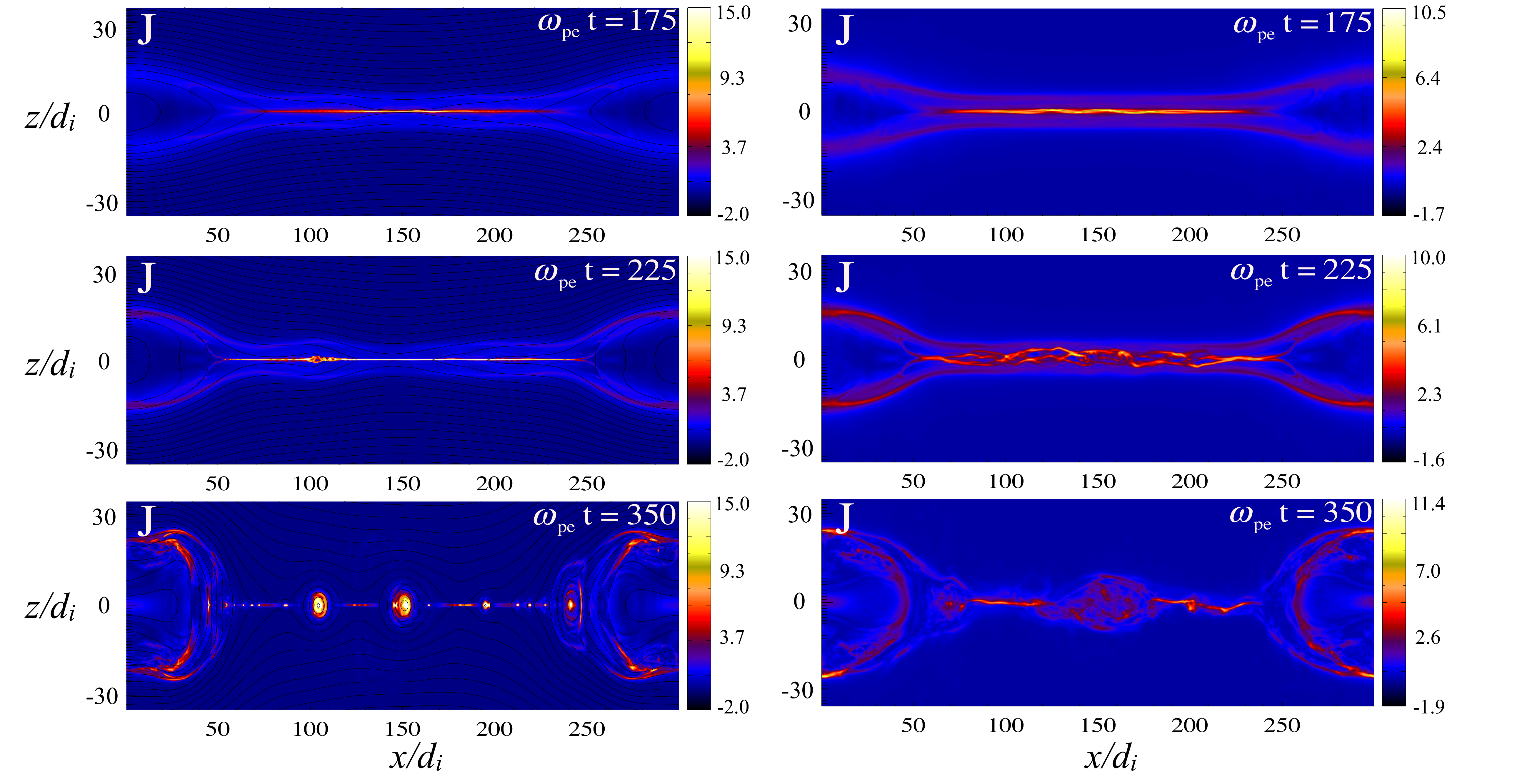}
\caption{Time evolution of 2D and 3D simulations with $\sigma =100$. Left: Color-coded current
density at $\omega_{pe}t = 175, \, 225$, and $350$, 
respectively. Right:  2D cut of current density from the 3D simulation at $\omega_{pe}t = 175, \, 225 $,  and $350$, respectively.}
\end{center}
\end{figure}

The evolution of the reconnection layer in the 3D 
simulation is illustrated in Figure 2, which shows
several snapshots of volume rendering of 
the current magnitude. Similar to the 2D case, initially the layer thins 
down under the perturbation that is uniform in the $y$ direction. However, 
the tearing instability and kink instability rapidly 
grow and the reconnection layer becomes strongly
turbulent. Throughout the simulation,
small scale ($\sim d_e$) kinked 
flux ropes are generated, and these quickly merge
into large ropes. The scale 
of the small scale ropes is similar to that in the 2D 
simulations.  The turbulence is fully developed to a power
spectrum with a clear sign of inertial range that has an 
index ``$-2$'' \citep{Guo2015}.

\begin{figure}
\begin{center}
\includegraphics[width=\textwidth]{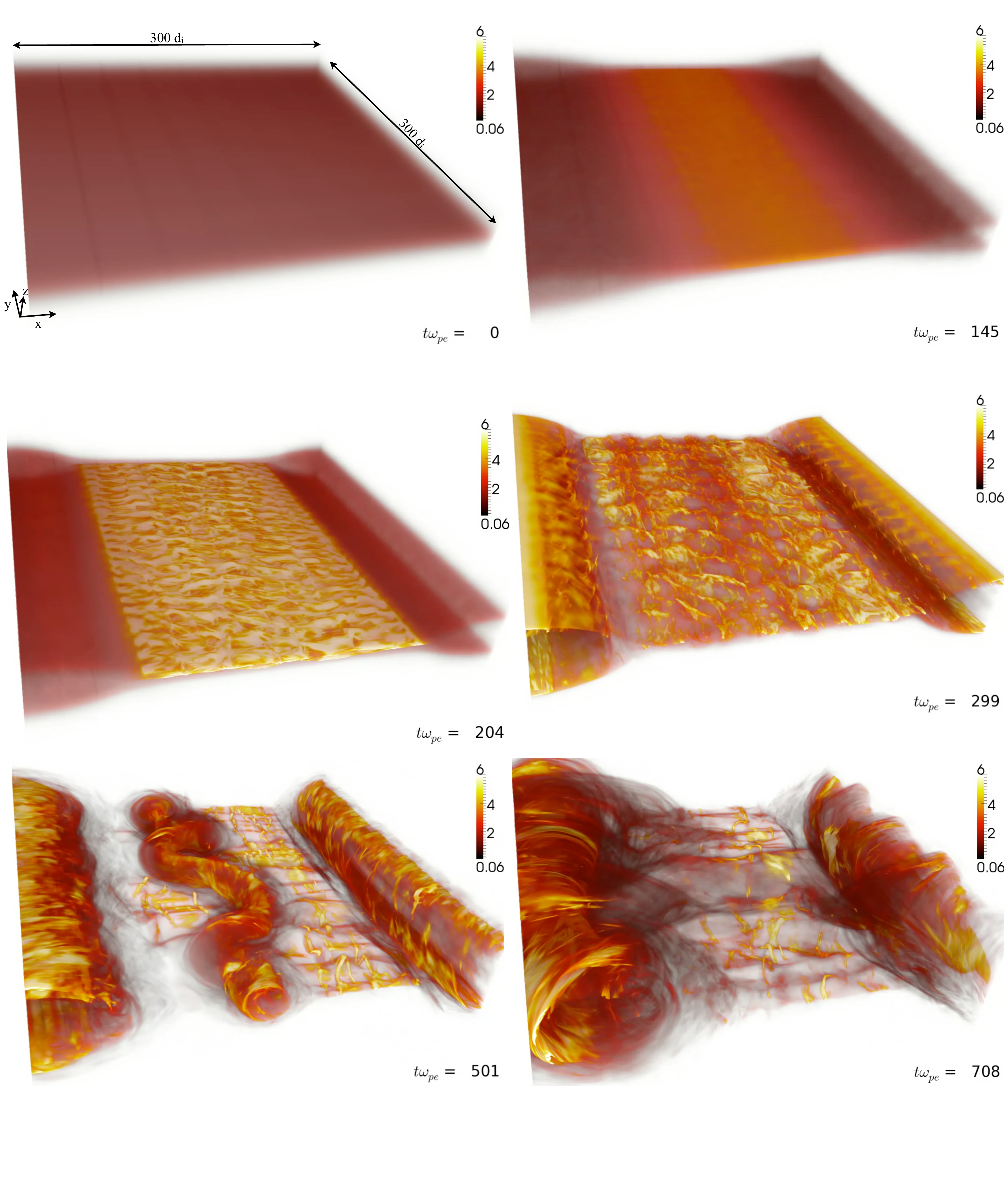}
\caption{Volume rendering of the current magnitude in the 3D simulations with $\sigma =100$ at different times.}
\end{center}
\end{figure}

Figure 3 shows the color-coded diagrams of (a) the bulk 
momentum in the $x$ direction $P_x = \Gamma v_x/c$, (b) the 
bulk momentum in the $z$ direction $P_z = \Gamma v_z/c$, 
and (c) the bulk Lorentz factor $\Gamma$. We find that the 
relativistic outflow can be generated in the reconnection 
layer. For higher $\sigma$, stronger bulk gamma can be 
found in the simulation \citep{Guo2015}. It has been shown 
that the reconnection rate and inflow outflow speeds are 
similar for Harris and force-free current sheet 
\citep{Liu2015}. The relativistic bulk motions may have a 
strong implication to the astrophysical high-energy radiation 
\citep{Giannios2009,Deng2015,Kagan2016}

\begin{figure}
\begin{center}
\includegraphics[width=0.8\textwidth]{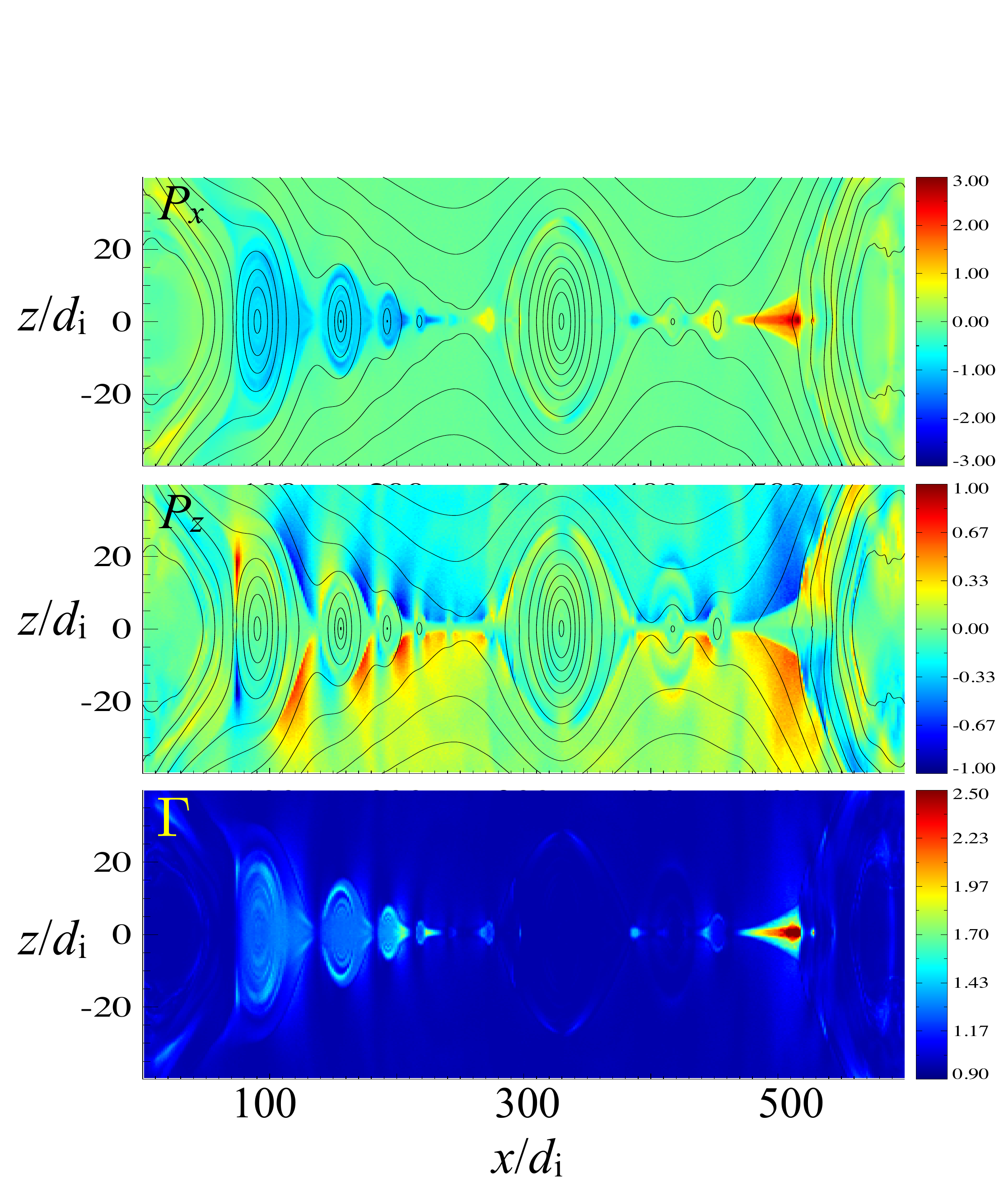}
\caption{The relativistic flows in the reconnection layer with $\sigma = 100$.
Top panel: the bulk momentum in the x-direction $P_x = \Gamma v_x/c$, Middle panel: 
the bulk momentum in the z-direction $P_z = \Gamma v_z/c$, Bottom panel: the bulk Lorentz 
factor $\Gamma$.
}
\end{center}
\end{figure}

Figure 4 shows the final energy spectra for $\sigma=100$ with different initial temperatures $T_e = 3.$, $1.0$, $0.3$, and $0.1 m_ec^2$, respectively. While for high temperature case the 
spectral index is close to $p=2$, for lower initial temperatures the energy spectra are harder and the spectral index approaches $p=1$. This shows that as the ratio between the magnetic energy
and the plasma energy increases, the spectral index becomes
smaller. 

\begin{figure}
\begin{center}
\includegraphics[width=0.8\textwidth]{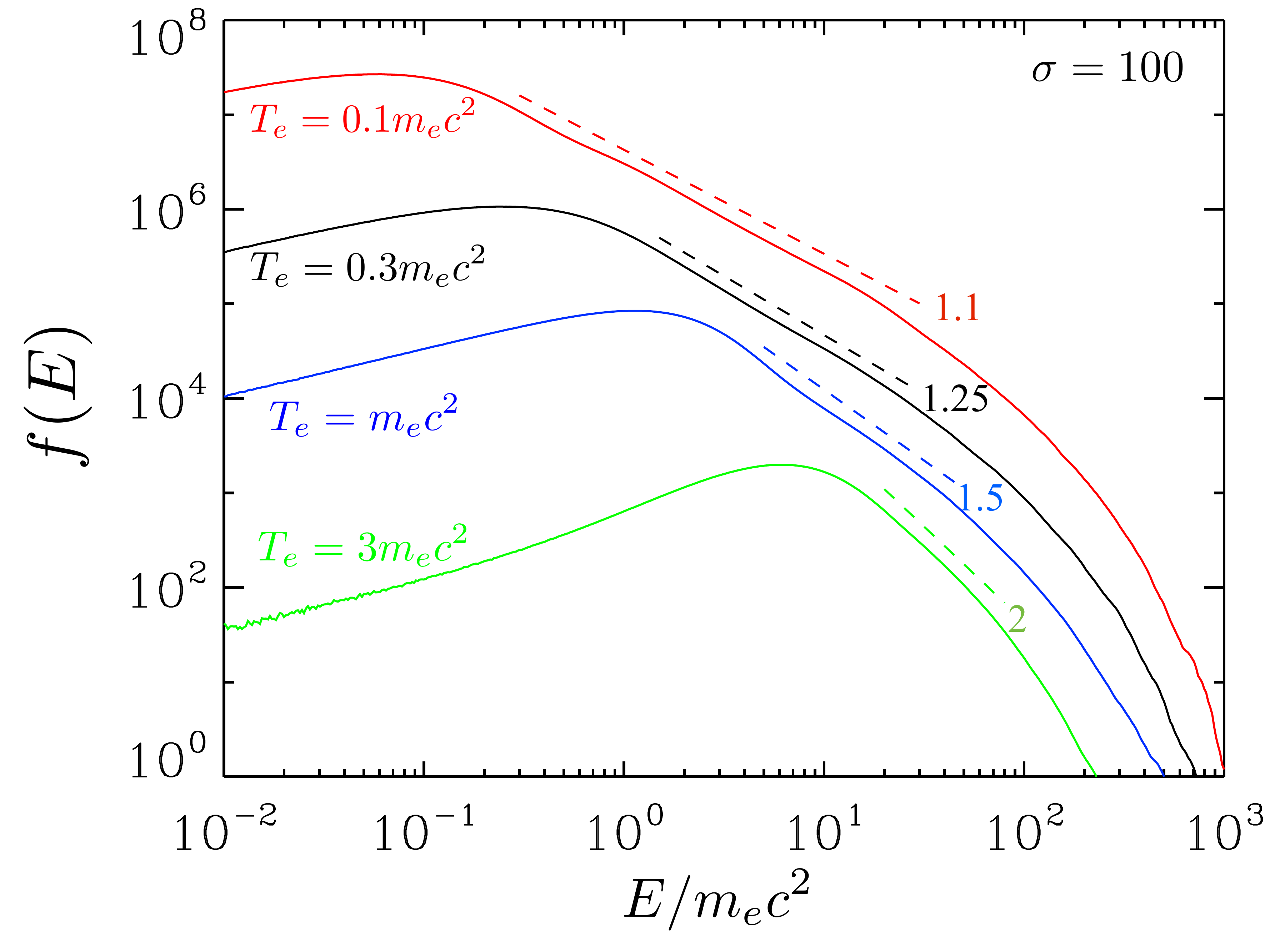}
\caption{The final energy spectra for cases with $\sigma=100$ and different initial thermal temperatures $T_i = T_e = 3.$, $1.0$, $0.3$, $0.1 m_ec^2$.
}
\end{center}
\end{figure}

Figure 5 shows the energy spectra for $\sigma=10$ with different initial temperatures $T_e = 1.0$, $0.3$, $0.1$, $0.03 m_ec^2$, respectively. While for high initial temperatures the energization is not significant deviated
from a thermal distribution, the cases with lower initial
temperatures show a $p \sim 1 $ energy spectrum.
The result is similar for the case with $\sigma=1$. Figure 6 shows the energy spectra for $\sigma=1$ with different initial temperatures $T_e = 0.1$, $0.03$, $0.01$, $0.003 m_ec^2$. While for high initial temperatures the energization is not significant deviated
from a thermal distribution, the cases with lower initial
temperatures show an overall $p \sim 1 $ energy spectrum. Therefore the generation of the nonthermal population of energetic particles appears to depend on the plasma $\beta$. As the plasma $\beta$ decreases, the released magnetic energy exceeds the initial plasma energy, which leads to a nonthermal energization.
\begin{figure}
\begin{center}
\includegraphics[width=0.8\textwidth]{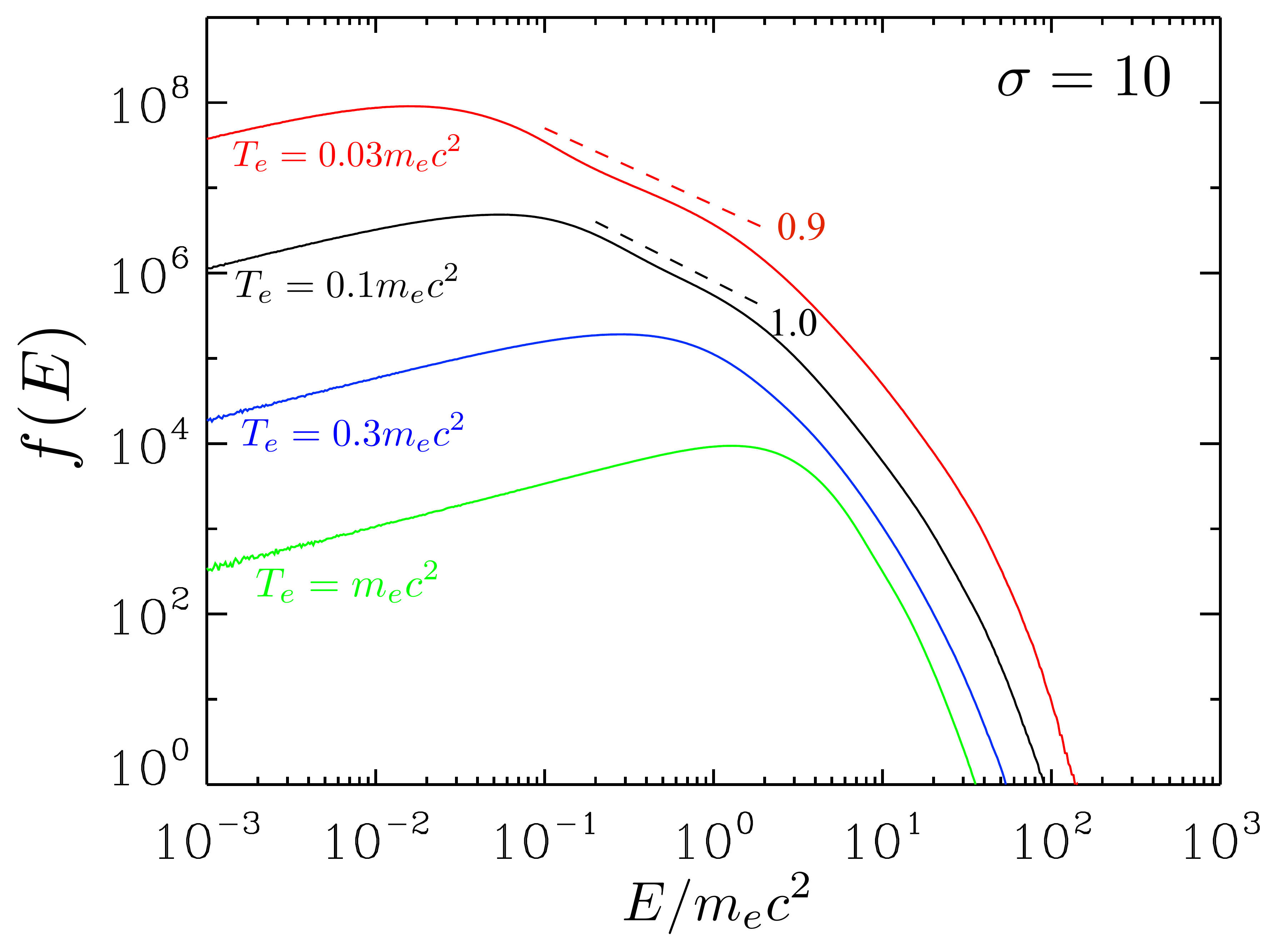}
\caption{The final energy spectra for cases with $\sigma=10$ and different initial thermal temperatures $T_i = T_e = 1.0$, $0.36$, $0.1$, $0.03 m_ec^2$.
}
\end{center}
\end{figure}

\begin{figure}
\begin{center}
\includegraphics[width=0.8\textwidth]{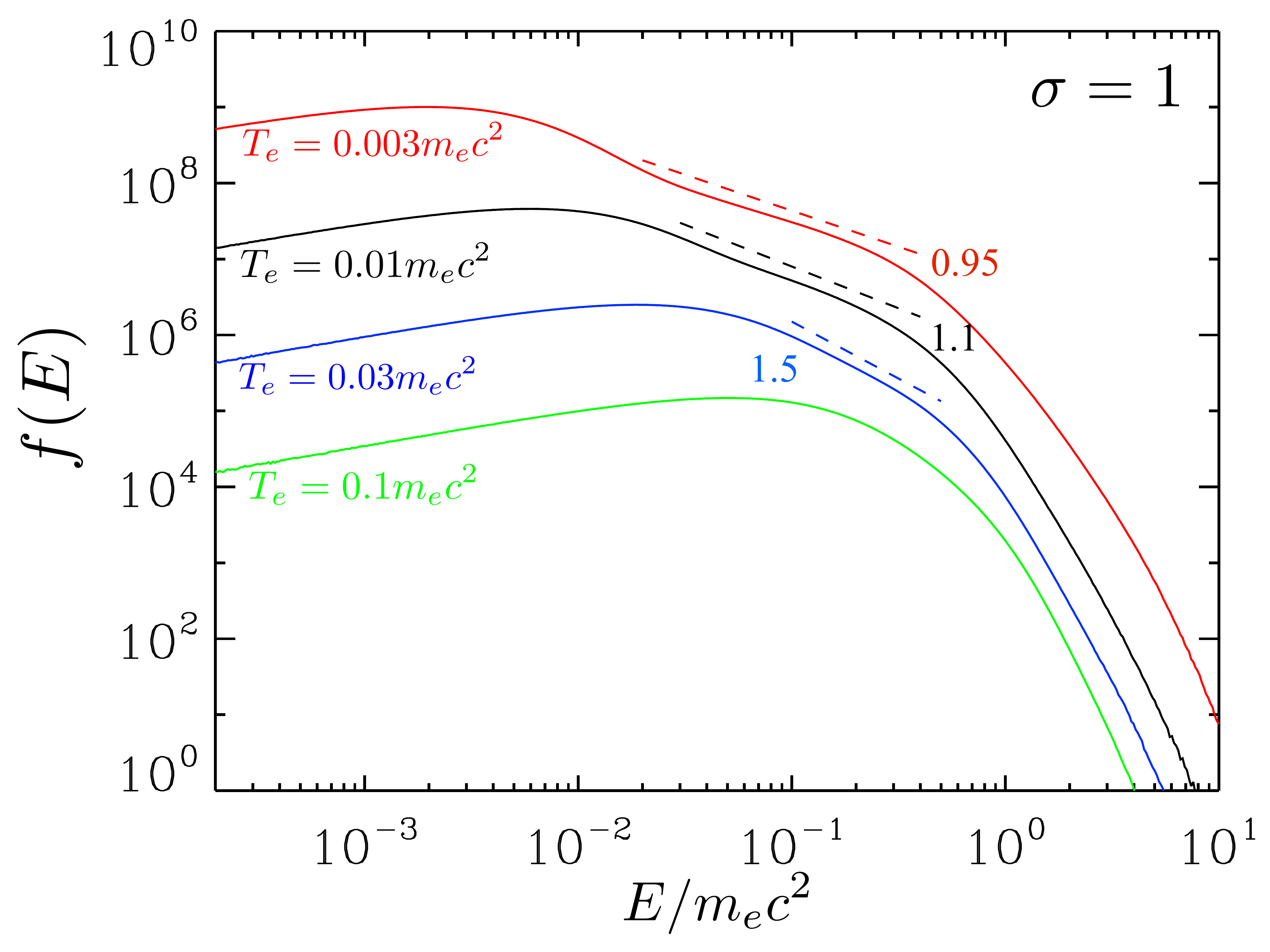}
\caption{The final energy spectra for cases with $\sigma=1$ and different initial thermal temperatures $T_e = 0.1$, $0.036$, $0.01$, $0.0036 m_ec^2$.
}
\end{center}
\end{figure}

\section{Outstanding issues and concluding remark}

The dissipation of magnetic field and particle
energization in the magnetically dominated systems 
is of strong interest in high energy astrophysics. 
In this study, we have briefly reviewed recent 
progress and further studied the nonthermal particle 
acceleration. The primary new results of the paper is 
that the initial temperature plays a role in determining
the spectral index of the nonthermal spectrum. While several 
earlier papers have concluded that the spectral index 
is smaller for higher $\sigma$, our simulations show 
that the spectral index approaches $p = 1$ for 
sufficient low plasma $\beta$. While so far the results
in general consistent with our analytical prediction 
in the earlier papers \citep{Guo2014,Guo2015}, it will be
interesting to study the case with lower $\beta$ when
we are able to reduce the numerical noise that may
cause artificial numerical heating.
These new results need to be
considered in interpreting the acceleration mechanisms 
from the PIC simulations. We also note that there are
a number of other issues that cause uncertainties in the 
reconnection acceleration theory. Below we outline several
issues that need to be addressed in the future.

\subsection{The dominant acceleration mechanism and power-law formation mechanism}
It should be noted that although multiple papers have 
demonstrated efficient nonthermal energization and 
the formation of power-law distribution using 
PIC simulations, the dominant acceleration mechanism and 
the formation mechanism for the power-law distributions 
have not reached a consensus (see Section 2 for a 
discussion). Two main possibilities discussed in the 
literature are direct acceleration by the nonideal 
electric field in the diffusion region 
\citep{Sironi2014,Bessho2012} and Fermi-like acceleration 
in the electric field induced by the motion of the 
reconnection driven flows \citep{Guo2014,Guo2015}.
Further efforts are required to distinguish the relative
importance of the two (or other) mechanisms and their roles
in the formation of power-law distribution and determining
the final spectral index.

\subsection{The effect of 3D physics and MHD turbulence}
Because of the level of computational cost, most of the 
kinetic studies of magnetic reconnection have been 
focusing on two-dimensional studies. 
There have been only a few 3D kinetic simulations 
of sufficient scale to allow a realistic interaction 
between various modes.
For example, it has been shown that the oblique tearing modes 
and kink modes develop 
and interact each other, leading to a turbulent reconnection 
layer \citep{Yin2008,Daughton2011,Liu2013,Guo2015}. However, those simulations have found about the same reconnection rate compare to 2D studies, indicating the 3D effects do not significantly alter the reconnection rate, although what determines the reconnection rate found in kinetic simulations is still a controversial topic. 
While early simulations show that kink instability
may prohibit the nonthermal acceleration \citep{Zenitani2008},
recent large scale simulations have shown that nonthermal 
acceleration can still develop despite the growth of the kink 
instability \citep{Sironi2014,Guo2014,Guo2015}. It will be 
interesting to analyze the effects of 3D physics to different
acceleration mechanisms for the nonthermal acceleration.

A closely related topic is the influence of turbulence on 
reconnection. The effects of MHD turbulence on the 
reconnection physics and the 
acceleration of particles have not been fully understood. Several
numerical studies have shown that magnetic turbulence can develop
from a three-dimensional reconnection layer, but the evidence
that the turbulence has strong effect on reconnection physics is still missing. It will also be interesting to study if the 
self-excited or externally driven turbulence will significantly
change the mechanism for nonthermal particle acceleration.

\subsection{Effects that lead to a steeper spectrum}

In agreement with other recent papers 
\citep{Sironi2014,Guo2014,Melzani2014b,Guo2015,Guo2016,Li2015,Werner2016}, this work shows that the spectral index in 
simulation is often much harder than commonly 
observed in space and inferred from astrophysical 
emissions. Although there is some observational
evidence in support of the 
hard spectrum \citep[e.g.,][]{Hayashida2015}, 
the power-law 
index predicted by the PIC simulation is systematically 
harder than most observations. More seriously,
for a power-law spectrum with spectral index $p<2$, the total
energy contained in the distribution quickly increases with
particle energy. This limits the maximum energy in the power-law predicted from the available magnetic energy.
We have 
analytically shown that allowing particle escape 
from the reconnection region
will produce a steeper spectrum \citep{Guo2014,Guo2015}. 
However, most of the 
kinetic simulations so far have used periodic boundary 
conditions. Nevertheless, these recent results suggest
that it is important to consider the effects that 
can lead to a softer spectrum such as open boundary 
simulations in the future. 

\section*{Acknowledgement}
This work is supported by the DOE through the LDRD program 
at LANL, DOE/OFES support to LANL in collaboration with 
CMSO, and by NASA through the Heliospheric Theory Program. 
X.L. is supported by NASA Headquarters under the NASA Earth and Space Science Fellowship Program -- Grant NNX13AM30H. The research is part of the Blue Waters sustained-petascale computing project, which is supported by the NSF (Grand No. OCI 07-25070) and the state of Illinois. Additional simulations 
were performed with LANL institutional computing. 


\end{document}